# IDENTIFYING RANSOMWARE ACTORS IN THE BITCOIN NETWORK

Siddhartha Dalal, Zihe Wang and Siddhanth Sabharwal

Columbia University, New York, USA
sd2803, zw2624, ss5689 @columbia.edu

## ABSTRACT

*Due to the pseudo-anonymity of the Bitcoin network, users can hide behind their bitcoin addresses that can be generated in unlimited quantity, on the fly, without any formal links between them. Thus, it is being used for payment transfer by the actors involved in ransomware and other illegal activities. The other activity we consider is related to gambling since gambling is often used for transferring illegal funds. The question addressed here is that given temporally limited graphs of Bitcoin transactions, to what extent can one identify common patterns associated with these fraudulent activities and apply them to find other ransomware actors. The problem is rather complex, given that thousands of addresses can belong to the same actor without any obvious links between them and any common pattern of behavior. The main contribution of this paper is to introduce and apply new algorithms for local clustering and supervised graph machine learning for identifying malicious actors. We show that very local subgraphs of the known such actors are sufficient to differentiate between ransomware, random and gambling actors with 85% prediction accuracy on the test data set.*

## 1. INTRODUCTION

Ransomware is a class of malicious software that, when installed on a computer, prevents a user from accessing the computer usually through unbreakable encryption until a ransom is paid to the attacker. In this type of attack, cybercriminals profit from the value victims assign to their locked data and their willingness to pay a fee to regain access to them. Bitcoin is a popular cryptocurrency used by ransomware actors to get ransom as it shields a person's personal identity by allowing them to transact using a Bitcoin address. Further, a bitcoin account holder (i.e., an actor) can create and hide behind multiple bitcoin addresses on the fly. Many fraudulent actors exploit this Bitcoin's pseudo-anonymity for their nefarious purposes. Prominent recent ransomware examples are Locky, SamSam, or WannaCry. As reported by Paquet-Clouston, et al [1], the latter infected up to 300,000 victims in 150 countries and that their lower bound estimate of the amount of bitcoin involved in ransomware transactions between 2013 to 2017 is more than 22,967.94 bitcoins amounting to over a billion dollars at the current exchange rate of 1 BTC = $46,491.11 [02/2021].

The goal of this investigation is to develop systematic ways to identify fraudulent actors in the Bitcoin network through graph classification. This is done by collecting data from multiple public sources on known ransomware addresses reported by their victims. These are used to generate connected transaction graphs in a limited time window. Since an *actor* (i.e., an account holder) can have many addresses, we identify the bitcoin addresses belonging to the same actor by a new method of local clustering, create features from subgraphs of Actor-to-Transaction bipartite graphs and identify other suspect ransomware actors using supervised machine learning. Figure 1 depicts the overall pipeline. Within the limitations discussed in this paper, we show that we can identify ransomware and gambling actors compared to a random account with accuracy of around 85% on the test dataset.

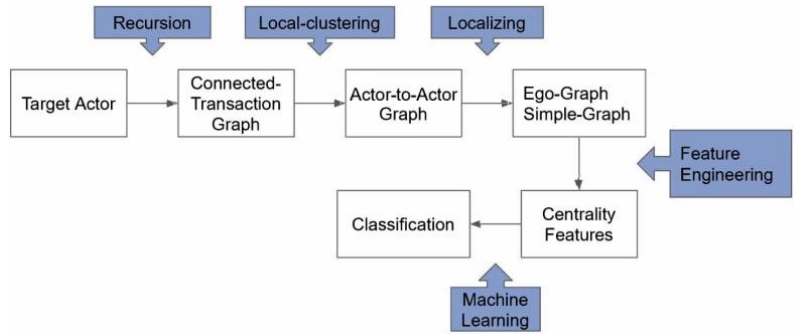

Figure 1. *Pipeline of the Approach: Acquisition, creation, wrangling and classification of data. Transformations are indicated in blue boxes.*

Specifically, Section 2 summarizes some of the previous work. Section 3 discusses the generation of data by web scraping Bitcoin addresses that have been tagged as being a part of a scam by other users in a number of public forums. Section 4 discusses and develops the corresponding temporally limited transaction graphs and the corresponding local clustering strategies. Section 5 proposes ego-graphs generated for ransomware, gambling and random actors for analysis along with several graph centrality metrics as features for supervised learning. Some data analysis is described in Section 6. Strategy for the supervised learning is described in Section 7 with the results of our analysis given in Section 8. Section 9 described limitations of our study and suggests future directions, The final section, Section 10, gives conclusion along with a brief discussion.

## 2. PREVIOUS WORK

Previous work on this topic can be divided in two parts. The first deals with how to cluster or link various addresses owned by a single actor, and the second discusses ransomware payments.

Many of the so-called behavioral address clustering algorithms are heuristics based. Specifically, Meiklejohn et al. [2] proposed two address-linking heuristics, namely (1) inputs spent to the same transaction are controlled by the same actor and (2) change addresses are not reused. *Change addresses* are commonly used by an account holder to preserve anonymity by creating multiple addresses and transferring bitcoins between those addresses. Indeed, it is considered as a good practice (Nakomoto [3]) to create a new address for transferring the remainder of bitcoins to this newly created address after transferring money to another actor. Harrigan et al [4] highlight that the clustering methods critically depend on the address reuse behavior. Goldfeder et al [5] extends the heuristics to cover CoinJoin transactions. However, Kalodner et al [6]) found that using these set of heuristics resulted in one super cluster of with 139 million addresses and many clusters with over 20,000 addresses. This happens mainly because when taking transitive closure of clusters, the errors are propagated across the entire bitcoin blockchain.

Another heuristic approach focuses on tracking IP-addresses, see Biryukov et al [7]. We do not pursue this line of inquiry since the Bitcoin Blockchain doesn't store the IP addresses; it has to be obtained by getting the log information from e-wallets or mempool. Further, these approaches have a low success rate, from 11% to 60%, as described by Biryukov et al.

While we follow behavioral clustering ideas, we modify them in a number of ways including local clustering. Specifics of algorithms are described in Section 4.

There have also been a number of attempts at using supervised learning to try to classify different categories of actors. For example, Harlev et al [8] uses clustering provided by Chainalysis to classify 434 clusters in different categories that include ransomware, Exchanges, mining pools, gambling, etc. They use also various machine learning algorithms including decision trees, boosting, random forests, etc. They report classification accuracy of 75% and higher. However, since they used clustering provided by Chainalysis, it would not be possible to identify new ransomware addresses without clustering. Further, since their reported results seem to be based on training data with no cross-validation or test data, their results are likely to be highly optimistic and overfitting (Hastie et al [9]).

Jordan et al [10] also consider a similar problem using graph motifs for classifying Exchanges, Services, Gambling, etc. They mention accuracy of around 90%. The novelty of their approach is the use of graph motifs to derive features for supervised machine learning. However, their analysis doesn't include ransomware actors. Further, their clustering algorithm uses only transitive closure of input addresses. They do not take into account CoinJoin, Coinbase or burn transactions. As mentioned in the clustering part earlier, this is likely to create many false positives. Further, again the accuracy reported seems to be based only on training data. Zola et al [11] also pursue motifs to do supervised learning. They improve results compared to a base model by using cascading classifiers, which uses cascades across 1, 2 and 3 motifs. Their reported results are impressive with accuracy score of up to 98%. However, again they do not consider ransomware actors, nor does their clustering takes into account CoinJoin, change of address or burn transactions.

An alternate approach based on deep learning has been proposed by Jung et al [12]. He proposes the use of Graphical Convolutional Neural Net Models (GCN). Besides being a black box, one of the major problems in using GCN is that each graph for GCN approach has to have the same number of nodes, which is not the case here. Further, the work did not cluster the addresses, and the reported accuracy of 72% is based on binary classification and only on training data. Our results are better.

Encouraged by Jordan et al [10] and Zola et al [11], we pursue a more detailed approach. Our contribution differ from them in a) our clustering algorithms are local and do take into account CoinJoin, change of address and burn transactions; b) we use ego-graphs instead of motifs - ego-graphs consider relationships between all the actors in the motif graphs, which include triangles, c) our data set includes the ransomware category; d) our features set is based on various explainable graph centrality metrics; finally e) our analysis is validated by using cross-validation as well as a separate test dataset.

## 3. CREATION AND WRANGLING OF DATA

### 3.1. Sources of Data

There were two key sources of data. First, sources that had addresses tagged as being associated with ransomware, i.e., for our "ransomware" class. Second, a source that had a comparison set of addresses that were not associated with ransomware, i.e. for our "random" and "gambling" classes. The random and gambling addresses are used as a comparison group for supervised machine learning.

Before analyzing the transaction pattern of these addresses, we needed to compile all the Bitcoin transaction data. It was downloaded from the Bitcoin Blockchain using Bitcoin Core [19] and we then accessed the raw data files which contain the validated transactions. Our study included data till July 2019 with 400,000,000 transactions and close to 40,000,000 addresses. The binary raw data was converted to a more human accessible format for analysis using BlockSci [6], which is an in-memory analytical database that allows for fast exploration over blocks and transactions due to their sequential, append-only generation process. With this data we had access to the entire transaction history for all addresses.

The first source consisted of creating a database of addresses of known ransomware actors. People who have been victims or approached for ransom, often publish the bitcoin address where bitcoins were asked to be sent as ransomware. Bitcoin WhosWho [21] and Bitcoin Abuse [22] were the two main sites which have such user-submitted addresses. Besides these two sources, we gathered information from the previously published literature [14] and law enforcement published actions (e.g., SEC). Much of the work involved going to these websites and scrapping the relevant information. All the addresses were collected in 2019.

The second source was Wallet Explorer [20]. It is a website that allows you to view the blocks and the individual transactions inside that block. From there one can also view the addresses and amounts involved in the transaction. The creator of the site Ales Janda wanted to associate addresses with actors. To do this he registered with a variety of businesses that accepted Bitcoin and transacted with those businesses. He then followed the Bitcoins he sent and catalogued which "wallet bitcoins were merged with, or from which wallet it was withdrawn." [20]. He then categorized each one of the businesses he catalogued addresses for into one of five categories: Exchanges, Pools, Services/other, Gambling, and Old/historic. The addresses were collected from Wallet Explorer in May 2019.

The Old/historic category contains many defunct businesses that early Bitcoin users used. The addresses that are associated with these businesses have been catalogued and the transactions that legitimate users had with these businesses have been web-scraped. Transactions associated with this category are what we call our "Random" class. The reason we call it as such is simply because the other businesses Janda compiled were all tagged with a specific category and these ones were across many other miscellaneous categories.

The gambling addresses we have collected are all of the Bitcoin addresses that have sent or received money from any of the associated gambling websites like CoinGaming, PocketDice, and BitcoinPokerTables but are not directly tagged to those websites. These websites are primarily designed so users can gamble using Bitcoin, however they sometimes have the added consequence of allowing money laundering to occur as users can "clean" their stolen Bitcoin into cash [13]. Transactions associated with these gambling addresses are what we call our "Gambling" class.

In total we collected 143, 498 and 216 ransomware, random and gambling addresses.

### 3.2. Graph Creation

For analysis, we need to extract relevant data of ransomware, gambling and random addresses from the Bitcoin Core. Bitcoin has three primary connected components: transactions, addresses and bitcoin transferred. From these, transactions-address bipartite graph can be created. Transactions are arranged in a set of sequentially linked blocks generated randomly approximately every 10 minutes (around 144 blocks per day). Each transaction has a set of input addresses, a set of output addresses and the amount of bitcoins transferred between them. There is also a transaction fee paid to the miner- an address that created the block.

However, there is no input for the Coinbase transactions, which are algorithmic transfers of bitcoins, one in each block to the corresponding miner.

### 3.1. Transaction Graphs

In the simplest terms, the transaction graph consists of a directed acyclic graph (dag), *T, A, W*, with transactions in the set *T* as nodes, input to output addresses in the set *A* as directed edges between transactions, and the bitcoin transferred in the set *W* as edge weights. Except for the transactions in the first block (the so-called genesis block) and Coinbase transactions, each transaction node is connected to multiple previous transactions as input nodes. A transaction node may not have an output node at a given time since there may be no transactions utilizing the output addresses of that transaction as input address so far. Specifically, a directed edge from node X to node Y means that an output address in X was an input address in Y and spent all of the Bitcoin they received in X in Y.

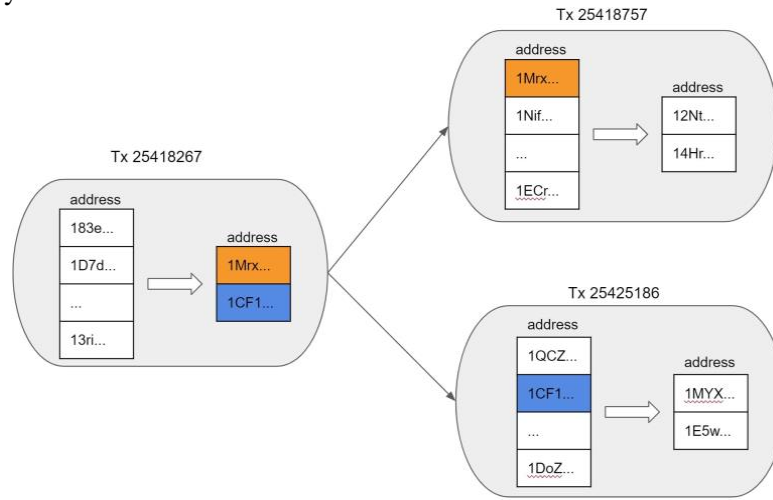

Figure 2. Anatomy of Transaction Graph: the transaction on left is connected by two different addresses (orange and blue) to two different transactions on right

Obviously, the transaction graph of the bitcoin transaction starts from the genesis block to the last block being considered. However, our interest lies in the behavior of the entity represented by a given address with the hope of identifying common patterns. Thus, we look at only the transactions involving an address under consideration (a random, gambling or ransomware address) and extend it iteratively to the transactions feeding the actor's transaction and the transactions being fed by the actor's transactions. Towards that end, given an actor $A$, we identify the first transaction $T_A$ involving that actor, and iteratively identify the transactions $T_p$ feeding to $T_A$ and transactions $T_f$ being fed by it.

This defines children-parents relationship. This process is followed iteratively taking transitive closure of all the children and parent transactions of $T_p$ and $T_f$ in the set of all transactions T. Though this limits the cardinality of the newly formed transaction set $\boldsymbol{T_A}$ corresponding to an address $A$, the beginning of the chain can still reach the genesis block or Coinbase transactions.

$\boldsymbol{T_A}$ can be further reduced for our problem. Based on looking at the behavior of a number of ransomware actors in their corresponding transaction graph $\boldsymbol{T_A}$, it was felt that just like many other criminals, the ransomware artist, after receiving the payment would rapidly make a succession of transactions to other addresses to make tracing difficult. Thus, it was decided to concentrate only on temporally local behavior of the address in $\pm n$ blocks. Specifically, for defining this local behavior, we restrict T$_A$ further from our recursion by using the following 3 rules.

1. Given a first transaction $T_A$ in $\boldsymbol{T_A}$ by the address $A$ under consideration, let the set $\boldsymbol{T_{A,n}}$ in $\boldsymbol{T_A}$ represent all the transactions in blocks $\pm n$ height away from the block containing the transaction $T_A$. For example, if transaction $T_A$ was in block 10,000 then the set $\boldsymbol{T_{A,n}}$ represents all the transactions between blocks 10,000 $\pm n$.
2. We further restrict $\boldsymbol{T_{A,n}}$ when the output side is an address belonging to an exchange or

gambling business as identified by Wallet Explorer. We do this because the children transactions links of the exchange node have many actors that have nothing to do with each other. The analogy is with a bank or a casino; if an actor was to deposit money in the bank or buy chips at the casino, we do not want to follow all the other actors who dealt with the bank or redeemed chips at the casino as they aren't necessarily linked to the actor of interest.

3. As an exception to the stopping criterion 1, we do not follow a Coinbase transaction backward, where a miner is awarded new bitcoins since there are no parent transactions.

Besides these criteria, we restricted $T_{A,n}$ further as stated below:

- *Non-standard Scripts*. There were several cases that Blocksci or various other explorers could not parse the address and would return a NaN, which can result in a situation that the output of source transaction or the input of the destination transaction or both are NaN. In this situation, in order to prevent the loss of information, we created dummy addresses to replace the NaN, unless we found an explorer which could parse the script. In that case we manually inserted the correct address.

- *Proof of Burn. OpReturn* transactions where an address may burn bitcoin to save a data item on the blockchain were assigned a string 'burn' to replace the NaN.

- *CoinJoin Transactions.* For the Actor-to-Actor graph creation, we need to identify mixing CoinJoin transactions. For more details on CoinJoin and other mixing transactions we refer to [15]. Given the new services like *Wasabi* [23] and *Samorai* [24], the older proposed identification rules do not work. We empirically modified rules used by BlockSci to tag CoinJoin transactions with the following rules.

    o If the transaction has less than 2 input or 3 output addresses, it is not a CoinJoin.

    o if the number of input addresses is smaller than the half of the number of the output addresses, the transaction is not a CoinJoin.

    o if the number of the output addresses is less than 6 and all output amounts are equal, the transaction is considered a CoinJoin.

    o if the number of the output addresses is more than 6, the transaction is considered CoinJoin if at least 5 output amounts are equal.

In summary, $T_{A,n}$ is the connected subgraph of the first transaction involving the actor $A$ within $n$ blocks either side with the exceptions described in the previous paragraphs.

For building the weighted transaction graphs, each of the edge of transaction graph had weight corresponding to transacted bitcoins. This required the information on the amount of input and output bitcoins and transaction fees in each transaction, maintaining the equilibrium - namely:

$$Input\_amount = output\_amount + transaction\ fees$$

As an example of $T_{A,n}$ Figure 3 depicts a sample of the directed transaction graph emanating from the Actor "12*HaVrpXkLr*2*UnkM f*6*X*9*b* on both sides. For ease of viewing all the self-loops have been removed, multiple edges have been collapsed and with no weights.

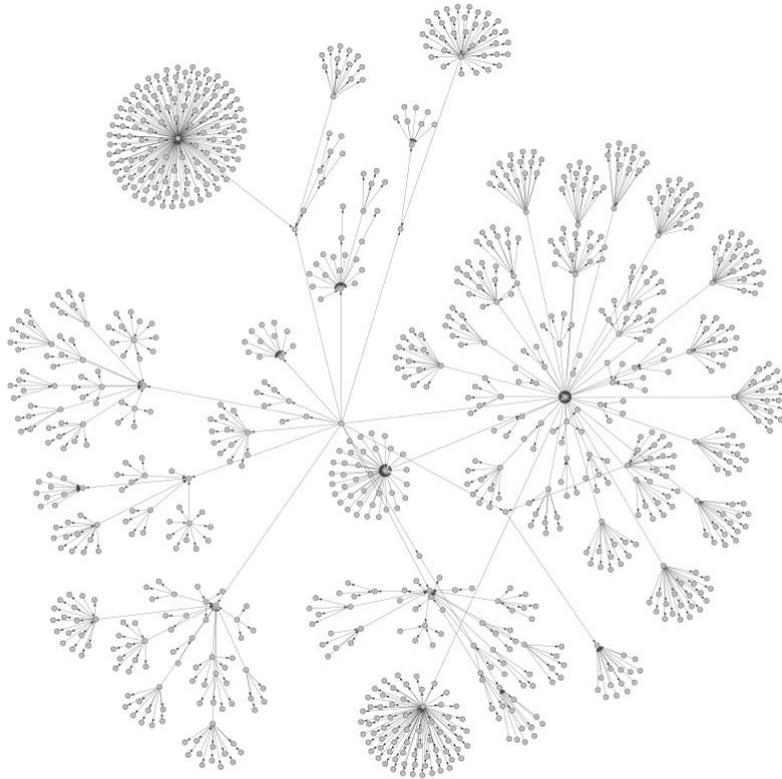

Figure 3. Actor 12HaVrpXkLr2UnkMf6X9bY11cuNrZUdUnV Transactions

## 4. CREATION OF LOCAL CLUSTERING AND ACTOR-TO-ACTOR GRAPHS

Since we are interested in the behavior of an actor, we need the corresponding actor-to-actor graph (not address-to-address graph). There are several difficulties with this. The main one lies with identification of the set of all addresses used by an actor. This is mainly, as mentioned earlier, due to bitcoin network allowing an account holder to create multiple bitcoin addresses on the fly. For simplicity we call the set of all addresses owned by an actor as an *entity* set and the corresponding graph is called an *entity graph* or an *Actor-to-Actor graph*.

As it is widely accepted, any address clustering scheme is imperfect, and ground truth is difficult to obtain on a large scale, since it requires interacting with service providers. Many other heuristics are possible, including those that account for the behavior of specific wallets. In our experiment with the data going to July 2019, when applying such heuristics to the entire blockchain, we got one super cluster containing more than 90% of addresses. This is primarily due to tumblers (the services which mix bitcoins) and CoinJoin kinds of transactions where multiple parties combine their transactions to preserve their anonymity. This is compounded by misattribution of the change of address.

We considered several modifications to the basic logic of behavioral clustering. But, when applied globally, all of them have exceptions which make a large number of false unions resulting in large clusters due to transitive closures. To limit potential for wrong clusters which gets propagated across the entire bitcoin blockchain, we adopt a different strategy of creating local clusters since our objective is mainly to identify scam artists who try to move bitcoin in a short period of time soon after starting their ransomware related scam. As discussed by Kharraz et al [16], just like any other crimes, ransomware artists move ransomware payments as quickly as possible. Thus, we decided to apply the clustering algorithm discussed earlier only locally within the temporal limit of $n=\pm 144$ blocks; basically within $\pm 1$ day.

We also apply somewhat different logic to CoinJoin and other transactions as described in the previous section. We may still have some false positives, but we will never have super clusters due to transitive closure. Any large clusters effects will be felt only within a particular local graph and not globally.

Specifically, we use the following rules:

1. Inputs spent to the same transaction are controlled by the same actor, thus, the entity set is the union of all those addresses.

2. If there is only one change address, identified by it never being used prior to the current transaction, consider it as a part of the input address set.

3. Exceptions to the rules 1 and 2 is when a transaction is identified as CoinJoin. In that case do not take a union.

The pseudo-code for the clustering process is given below.

---
**Algorithm 1 Generate Local Cluster**

**for** all transactions in the graph **do**
   **if** this is a CoinJoin transaction **then**
      assign each input and output addresses as separate clusters.
   **else**
      **if** there is only one new address in output addresses of this transaction
         assign all input addresses and output addresses as one cluster.
      **else**
         assign all input addresses as one cluster;
         assign each input and output addresses as separate clusters.
      **end if**
   **end if**
**end for**
After collected all address-cluster mapping, merge the mapping till there is only one cluster for each address

---

For the weighted graph analysis, we need to find the bitcoin transfer between addresses. Since the bitcoin transfer is defined between the sets of input addresses and output addresses, there is no exact way to allocate the amount between a given input address and an output address unless one of the sets has cardinality 1. We approximate the transfer by a proportional allocation rule. Namely, given a transaction with input addresses: $I_1, ..., I_x$; input amounts: $IA_1, ..., IA_x$; output addresses: $O_1, ..., O_y$; and output amounts: $OA_1, ..., OA_y$, the edge weight from $I_i$ to $O_j$ is computed by the following formula: $(IA_i/\Sigma IA_k)*OA_j$, which is further adjusted by the transaction fee. This is an approximation. For Actor-to-Actor graph, the weights are the sum of all individual weights of the corresponding addresses.

The development described so far allow us to form Actor-to-Actor weighted graphs. Some of these graphs after clustering had only a small number of nodes and were deleted. These were further sub-divided in training and test sets of 328 and 82 graphs (80−20%), respectively, for supervised learning.

## 5. GRAPHS AND CENTRALITY FEATURES

### 5.1. Subgraphs

Recall that the following the logic of the last sections, we consider the locally clustered *Actor-to-Actor* graph from the connected transaction graph $T_{A,1}$ for an actor $A$ within $\pm 144$ blocks ($\pm 1$ day). The subgraph of all addresses within $T_{A,1}$ is referred to simply as *whole* graph. For additional analysis, we take several different kinds of subgraphs. Specifically, since our primary interest lies in the actor under consideration, we created Ego subgraphs for the actor. *Ego graph* of order n of a node is the subgraph formed by the nodes that are within the neighborhood of order *n* of the node without considering the direction of the edges. Ego graphs are richer than standard motifs since they also consider relationships between neighbors. Further, another set of subgraphs, called *simple graphs* were obtained by removing loops of the nodes to itself, and collapsing multiple edges to one edge. These subgraphs are considered since it is expected that the actor's footprints would be most visible in its direct transaction with other nearby actors. For example, the ransomware actor's footprints would be most visible in its interactions with the victims and other nearby actors and co-conspirators. For further analysis we only considered ego1, ego2, ego3 and the corresponding versions of simple graphs.

### 5.2. Centrality features

For each of the subgraphs discussed above, we extracted a number of graph-based features:

i. Basic Statistics: # of Vertexes, # of Edges, Total bitcoins, Loops, Degree, Neighborhood size

ii. Centralities: Normalized Closeness, Betweenness, Page Rank, Cluster Coefficient, Coreness, Hub and Authority

Definitions of each can be found in **igraph** [26] with more details in article [17]. Some are overall graph parameters; the rest are restricted to the node of the actor under consideration. A number of variants of these were considered where it made sense including weighted, unweighted and directed. The creation of graph and extraction were all carried out by using *Python igraph* library [26].

The task of computing graphs and its features was computationally intensive. For efficiency reasons, we did not consider the graphs larger than one million unique addresses or more than 1∕2 million transactions. As mentioned earlier, *whole* graph with only a small number of nodes and the corresponding ego graphs were also removed, Finally, to balance the classes better a random sample of size 155 was taken from random graphs. An *80-20 split* between training and test data with stratification resulted in the training set of 328 *whole* graphs (124 random, 80 ransom,124 gambling), and the test set of 82 *whole* graphs (31 random, 20 ransom, 31 gambling).

## 6. EXPLORATORY DATA ANALYSIS

In this section, we describe an exploratory analysis on some of the features generated in section 5. Here, we used the 'boxen plot' which centers a distribution at its median line; each successive level outward contains half of the remaining data until it reaches to the outlier level. For details, see *seaborn* [27].

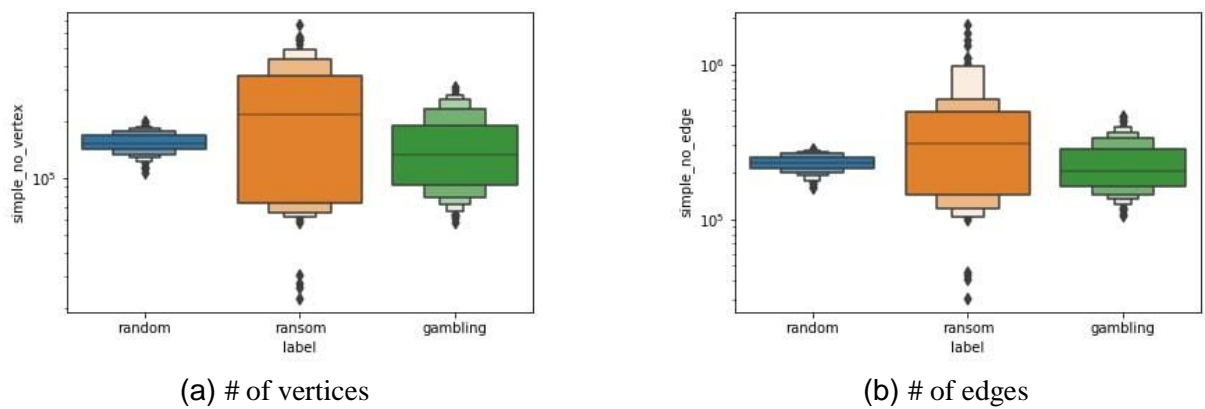

(a) # of vertices  (b) # of edges

Figure 4. whole-simple graph

Figure 4 shows the number of vertices and edges in the whole-simple Actor-to-Actor graphs. *Recall that the whole graph is based on recursion of all connected transactions associated within*

*2 days of the actor*. Thus, these graphs could be skinnier than the graphs over two days depending upon the level of connections of the actor. We can see that for 'random' and 'gambling' graphs, the distribution does not differ a lot. However, there are many extreme values in 'ransom' graphs and it is flatter compared to other two. It reflects the nature of the 'ransomware' class where actors will try to obfuscate their transaction patterns through complicated laundering, which also reflects that the local clustering algorithm performs well.

For brevity, the rest of the analysis highlights only Ego-1-simple graphs for a few important features found in the Results section since they are the simplest graphs of other actors who are in close touch with the Actor. The analysis looks at the marginal distribution of the selected features across all the actors separated by ransomware, random and gambling categories.

PageRank, also known as Google Rank, is a way of measuring the importance of website pages. The assumption is that more important websites are likely to receive more links from other websites. For more definition, see *PageRank* [28]. As seen in Figure 5, the 'ransom' clusters tend to have a higher PageRank, which means it is likely to receive more transactions from other clusters. This makes senses when the ransomware attack is one wave after another and usually sent to lots of users within a short period of time. Also, PageRank of random actor on average seems to be lower than ransomware actors indicating that the ransomware actors are more often recipients of funds.

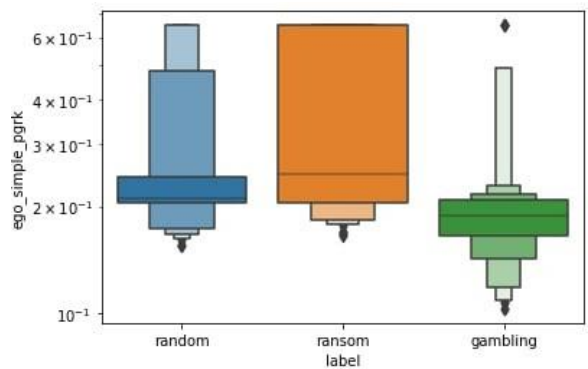

Figure 5. Page-Rank

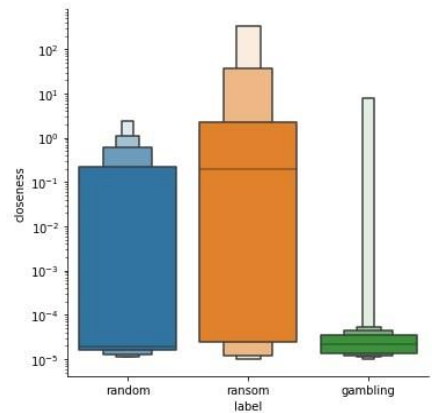

Figure 6. Closeness

The closeness centrality of a vertex measures how easily other vertices can be reached from it (or the other way: how easily it can be reached from the other vertices). For definition, see *closeness* [26], [31]. The weighted-IN closeness of ego-1 simple graphs is shown in Figure 6. The 'gambling' tends to have less centrality, which shows similar pattern as shown in Figure 5. This suggests that gambling actors are not closely connected to other accounts. They have many more outliers indicating that there are few very large gamblers and possibly the distribution is scale-free.

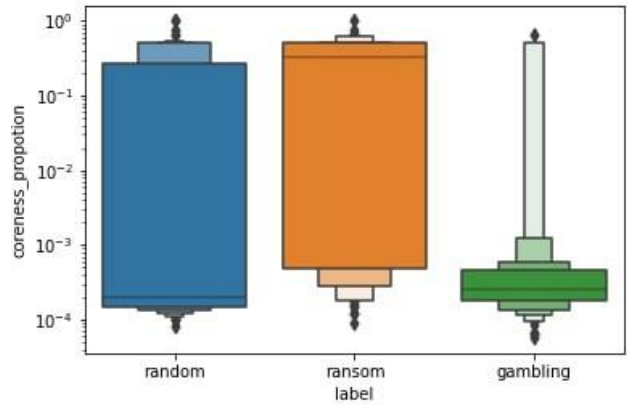

Figure 7. Coreness normalized by # of vertices

In Figure 7, we show the *coreness*(All). The k-core of graph is a maximal subgraph in which each vertex has at least degree k. The coreness of an Actor is k if it belongs to the k-core but not to the (k+1)-core. For the definition, see *coreness* [29]. The coreness across graph is normalized by the number of vertices since all the graphs have different number of vertices. From the figure, as before, it can be noticed that the gambling graphs have a relatively low coreness.

Finally, Figure 8 shows, as one would have expected that cluster coefficients for gambling class is much smaller and followed by ransomware class since both of those classes are directly involved in possibly criminal activities and they would minimize their interactions with actors which are more connected with each other.

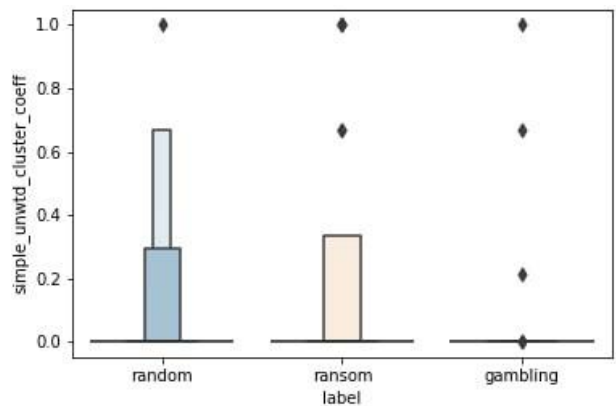

Figure 8. Unweighted Cluster Coefficient, Normal Scale

The comparative analysis of the marginal distributions of features reported so far suggests that different classes behave somewhat differently from each other. For example, gambling actors behavior is rather different than other actors in closeness, PageRank, cluster-coefficient and coreness. Further, the PageRank of the ransomware actors is higher. This analysis indicates that these features could be good candidates for any machine learning model.

## 7. MACHINE LEARNING ON ACTOR-ACTOR GRAPHS

For the purposes of supervised learning, the extracted whole graphs were divided in testing (20%) and training (80%) graphs stratified by their categories. Further, for each whole graph only the subgraphs of ego-graph 1, ego-graph 2, ego-graph 3 and their simple counterparts were extracted for analysis because of their proximity to the actor under consideration. Only subsets of features given in Table 1 were extracted from each of these graphs. The subset was obtained by keeping only one of each set of highly correlated features. The graphs and the corresponding number of features are shown in Table 1 below.

|  | ego3 | ego3-simple | ego2 | ego2-simple | ego1 | ego1-simple |
|---|---|---|---|---|---|---|
| # of features | 11 | 16 | 13 | 16 | 12 | 11 |

Table 1: Centrality Features Considered

### 7.1. Models

We consider supervised learning in three stages shown in Table 2.

| Learning | Initial | Intermediate | Final |
|---|---|---|---|
| Type | Multiple Classifiers | Stacking | Bagging |

Table 2: Modeling Strategy/Stages

In the Initial stage multiple classifiers were fitted to each of the 6 sub-graphs. Since each classifier has different strengths and weaknesses in different regions of the feature space, as an intermediate model, we used ensemble learning technique of Stacking [18] to improve. Specifically, the stacked model uses the predicted probabilities of each class by each classifier as features and predicts the probability of each class by using a simple model (logistic in our case). Even though we have 3 classes, since the probabilities add up to 1 for each of the six classifiers, there are 12 such linearly independent features. This process is depicted in Figure 9 with the six different base classifiers we used for creating the ensemble.

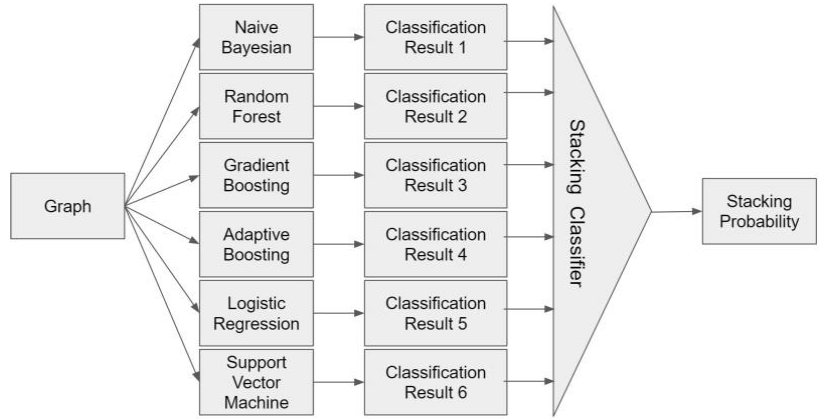

Figure 9. Stacking Model

In the final Stacking-Bagging stage, we combine the results across different subgraphs by creating a meta classifier (or called Final Classifier). It is a simple classifier (in our case logistic) that uses the probabilities of each class in the subgraph stacked models as the feature set and fuses them. This is analogous to bagging since we have 6 different data sets (subgraphs) each containing estimated probabilities of each class. Just like in stacking we will have 12 features for the six types of sub-graphs. The final attribution of the class is given to the class with highest probability by the meta-classifier. This process is depicted in Figure 10. We used the predicted probabilities from 6 graphs as new features and trained a final meta-classifier. We called this as 'stacking-bagging' model.

To motivate efficacy of the stacking-bagging model, consider the simple fusing by averaging the probabilities across 6 estimated probabilities for each data set. In that case, Mean Squared Error (MSE) = Bias^2 +Variance. Each component on the bias term is roughly the same constant since they are using the same type of estimators. The second term=average_variance/6+∑Covariances/6. In our case, since each estimated probability use different graphs, it is expected that the covariances would be relatively negligible. Thus, the MSE will be substantially less compared to non-fusing.

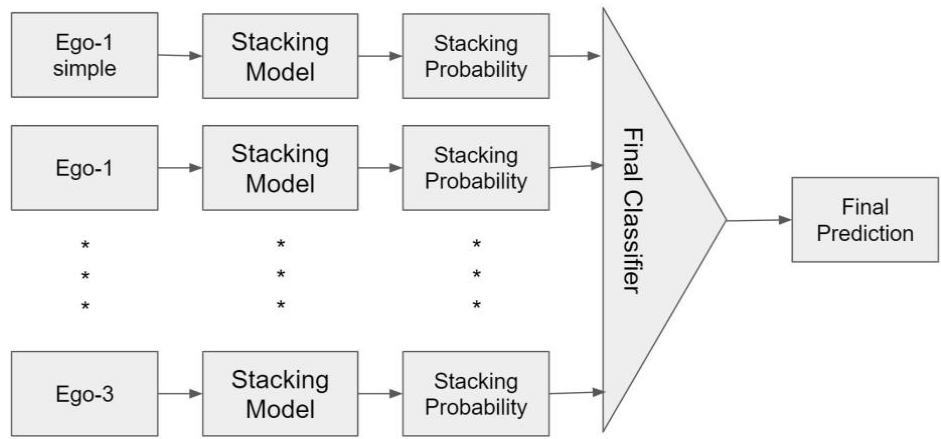

Figure 10. Stacking-Bagging Model

We used the cross-validation score on balanced-accuracy as our objective and finally ran our model on our test set.

## 7.2. Cross Validation and Efficacy Metrics

To implement the strategy outlined in 6.1, and to measure its efficacy we used the training data with cross validation for model selection. Specifically, when training the classifiers with grid-search and cross-validation, we used 5-folds with stratification on labels and 80% of the data for train-validation and 20% for testing. For the meta-classifier in the stacking model and for the stacking-bagging model, we used Logistic Regression.

Since we have a multi-label classification problem, we used balanced accuracy or simply Accuracy in tables), weighted precision (as Precision in tables), and weighted recall (as Recall in tables) as our evaluating metrics. We refer to them simply as accuracy, precision and recall. For definitions see *scikit-learn* [25].

## 8. RESULTS

Table 3 gives cross-validated accuracy on training data of the 6 base classifiers using the cross-validation on the training data. Gradient boosting and Random Forest models seem to outperform others with around 75% to 80% balanced accuracy. Taking a deep dive on feature importance, Figure on the left side of Table 3 shows the 11 features included in Random Forest model for the ego-1 simple graph. Besides the commonly used graph centrality features, it includes less used features like *coreness* and *cluster coefficient*.

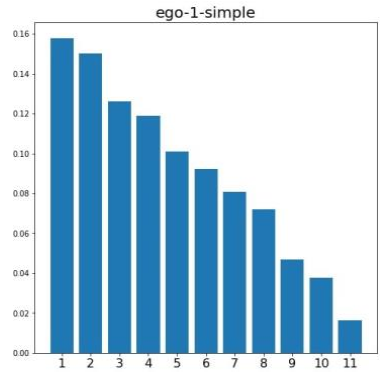

| Rank | Feature | Value |
|---|---|---|
| 1 | Closeness(wtd/out) | 0.158 |
| 2 | sum of weights | 0.150 |
| 3 | Closeness(uwtd/out) | 0.126 |
| 4 | # of vertices | 0.119 |
| 5 | Closeness(wtd/in) | 0.101 |
| 6 | cluster coefficient | 0.092 |
| 7 | Closeness(wtd/all) | 0.081 |
| 8 | Closeness(uwtd/in) | 0.072 |
| 9 | Coreness(all) | 0.047 |
| 10 | Authority | 0.038 |
| 11 | Coreness(IN) | 0.016 |

Table 3: Feature Importance of Ego-1-simple graph

|  | ego-1-s | ego-1 | ego-2-s | ego-2 | ego-3-s | ego-3 |
|---|---|---|---|---|---|---|
| Naive Bayesian | 0.4148 | 0.4324 | 0.5514 | 0.4034 | 0.5189 | 0.4047 |
| Random Forest | 0.7346 | 0.7924 | 0.7596 | 0.7966 | 0.7973 | 0.8171 |
| Gradient Boosting | 0.7255 | 0.7709 | 0.7456 | 0.7651 | 0.7829 | 0.8107 |
| Adaptive Boosting | 0.6911 | 0.6937 | 0.7111 | 0.7160 | 0.7444 | 0.7386 |
| Logistic Regression | 0.6162 | 0.6272 | 0.5862 | 0.6000 | 0.6101 | 0.5962 |
| SVM | 0.6126 | 0.6271 | 0.5997 | 0.6031 | 0.6298 | 0.5908 |

Table 4: Balanced Accuracy of Classifiers for each graph

Stacking these models produces cross-validated balanced accuracy between 96% and 99%, a substantial improvement. It is interesting to note that ego-simple graphs tend to outperform their corresponding ego graphs.

|  | Accuracy | Precision | Recall |
|---|---|---|---|
| Ego-1-simple | 0.9932 | 0.9942 | 0.9939 |
| Ego-1 | 0.9890 | 0.9910 | 0.9909 |
| Ego-2-simple | 0.9602 | 0.9661 | 0.9634 |
| Ego-2 | 0.9662 | 0.9690 | 0.9664 |
| Ego-3-simple | 0.9917 | 0.9941 | 0.9939 |
| Ego-3 | 0.9743 | 0.9735 | 0.9724 |

Table 5: Stacking Model: Performance by Cross-Validation

In the final stage, we build a bagging-stacking model on all six types of ego subgraphs leading to the cross-validated accuracy of 1 and 85% on the test set as shown in Table 6. The corresponding confusion matrix is given in Figure 11.

|  | Cross-Validation | Test |
|---|---|---|
| Accuracy | 1 | 0.8537 |
| Precision | 1 | 0.8566 |
| Recall | 1 | 0.8537 |

Table 6: Final Model Accuracy, Precision and Recall

As seen from the above table, the final model outperformed the stacking model as measured by cross-validated accuracy of 1. Figure 11 gives the corresponding confusion matrix for the final model on the test set. There is no systematic confusion evident in the confusion matrix.

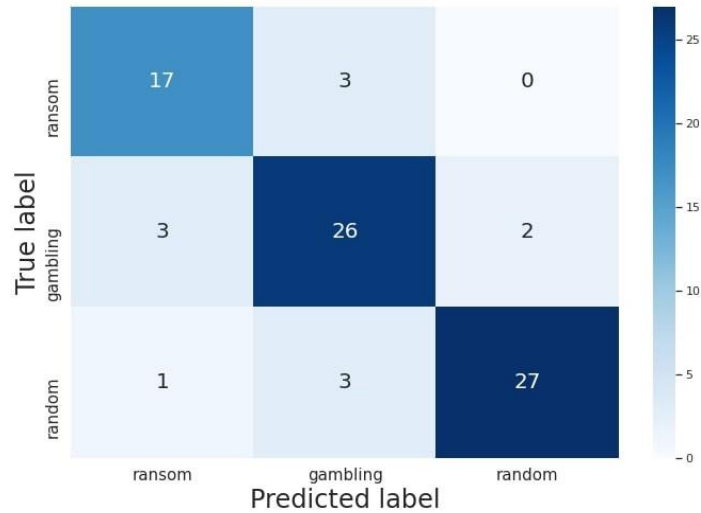

Figure 11. Confusion Matrix of the test data

## 9. LIMITATIONS AND FUTURE WORK

Though our results indicate high accuracy, further improvement should be possible by improving clustering algorithms, better ways to associate actors involved with non-standard scripts and Coinjoin transactions. Further, our data set is limited consisting of around 400 graphs, each with thousands of nodes. Further, our ground truth is based on what is user-reported. Getting more data which is more reliable would improve accuracy. Also, given that we are using stacking, it is hard to interpret the final model. More interpretative models using a different machine learning approach may be feasible. Finally, though we have not undertaken it here, it would be worthwhile trying to identify actors from TOR since they are also likely to be involved in illegal activities [30]. Another direction to explore would be to see how these techniques can be generalized to other alternative crypto-currencies.

## 10. CONCLUSIONS

This paper addresses the key question of how to identify miscreants who are involved in ransomware and in gambling compared to random actors. The problem is difficult due to the pseudo-anonymity of the Bitcoin network. Specifically, the question addressed here is that given temporally limited graphs of Bitcoin transactions, to what extent can one identify common patterns associated with these fraudulent activities and apply them to find other similar actors. The singular contributions of this paper include a) extraction and creation of transaction graphs associated with the miscreant actors, b) clustering all the nodes of such graphs in common entities controlling those accounts while taking into account different kinds of transactions, c) using supervised learning novel algorithms to create models based on actor to actor ego graphs that identify similar miscreants, d) validating the models on cross validated data with accuracy of 1 and on the test data set of around 85%.

## 11. ACKNOWLEDGEMENTS

We acknowledge Professor Lazaros Gallos, Jianqiong Zhan, Vatsal Randhar and Xiaoqi Wang for helpful discussions and contributions. Financial support from School of Professional Studies, Data Science Institute and Statistics Department at Columbia University is also gratefully acknowledged. Computing support was provided by Habanero High Performance Computing Cluster at Columbia University

## AUTHORS

**Siddhartha Dalal**

Siddhartha Dalal is Professor of Practice at Columbia University. He received his MBA and PhD from University of Rochester. Prior to joining Columbia, he was the Chief Data Scientist and Senior VP at AIG, CTO at RAND Corporation, VP of Research at Xerox and Chief Scientist and Executive Director at BellLabs/Bellcore.   He also advised the US Army and DoD on technologies. He has over 100 peer-reviewed publications, patents, and monographs covering the areas of risk analysis, medical informatics, Bayesian statistics and economics, image processing, and sensor networks. He has received several awards including from IEEE, ASA, and ASQ, notably for his work on Space Shuttle Challenger disaster and for managing software risks. The US Army has awarded him the Meritorious Civilian Service Medal.

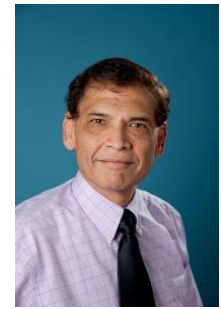

**Zihe Wang**

Zihe Wang is currently a research staff associate at Columbia University. He completed his Master's in data science from Columbia University in 2020 and his Bachelors in Statistics and Computer Science from University of Illinois at Urbana-Champaign in 2019.

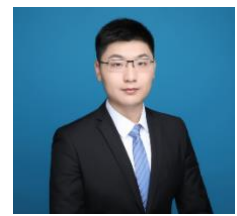

**Siddhanth Sabharwal**

Siddhanth Sabharwal is currently a first-year PhD student in the Statistics department at University of Illinois at Urbana-Champaign. He completed his Master's in Statistics from Columbia University in 2019 and his Bachelors in Statistics and Computer Science from University of California Davis in 2017.

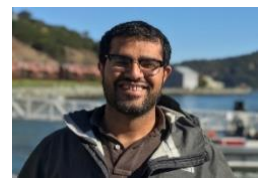